\documentclass[useAMS,usenatbib]{mn2e}

\usepackage[latin1]{inputenc}
\usepackage{multirow}
\usepackage{graphicx}
\usepackage{multicol}
\usepackage{float}
\usepackage{color}
\usepackage{footnote}
\usepackage{amssymb}
\usepackage[figuresright]{rotating}
\usepackage{amsmath}
\usepackage{url}
\usepackage{rotating}
\usepackage{caption}
\newcommand{\Mpc}{\rm\; Mpc}
\newcommand{\kpc}{\rm\; kpc}

\newcommand{\km}{\rm\; km}

\newcommand{\cm}{\rm\; cm}
\newcommand{\mum}{\hbox{$\rm\; \mu m\,$}}
\newcommand{\pix}{\rm\; pixel}
\newcommand{\Gyr}{\rm\; Gyr}

\newcommand{\s}{\rm\; s}

\newcommand{\ks}{\rm\; ks}

\newcommand{\GHz}{\rm\; GHz}

\newcommand{\rcool}{\hbox{$\thinspace r_\mathrm{cool}$}}
\newcommand{\tcool}{\hbox{$\thinspace t_\mathrm{cool}$}}
\newcommand{\Lcool}{\hbox{$\thinspace L_\mathrm{cool}$}}

\newcommand{\Msun}{\hbox{$\rm\thinspace M_{\odot}$}}

\newcommand{\keV}{\rm\; keV}

\newcommand{\erg}{\rm\; erg}

\newcommand{\ergps}{\hbox{$\erg\s^{-1}\,$}}

\newcommand{\kmps}{\hbox{$\km\s^{-1}\,$}}

\newcommand{\kmpspMpc}{\hbox{$\kmps\Mpc^{-1}\,$}}

\newcommand{\LIR}{\hbox{$\thinspace L_\mathrm{IR}$}}
\newcommand{\Lkin}{\hbox{$\thinspace L_\mathrm{kin}$}}
\newcommand{\LR}{\hbox{$\thinspace L_\mathrm{R}$}}

\newcommand{\Lx}{\hbox{$\thinspace L_\mathrm{X}$}}
\newcommand{\Lnuc}{\hbox{$\thinspace L_\mathrm{nuc}$}}

\newcommand{\Omm}{\hbox{$\rm\thinspace \Omega_{m}$}}
\newcommand{\OmL}{\hbox{$\rm\thinspace \Omega_{\Lambda}$}}

\newcommand{\pcmsq}{\hbox{$\cm^{-2}\,$}}

\title[Radiatively-inefficient cluster cores]{Investigating a sample of strong cool core, highly-luminous clusters with radiatively-inefficient nuclei}
\author[J. Hlavacek-Larrondo and A. C. Fabian]{J. Hlavacek-Larrondo$^{1}$\thanks{E-mail: juliehl@ast.cam.ac.uk} and A. C. Fabian$^{1}$\\
$^{1}$Institute of Astronomy, University of Cambridge, Madingley Road, Cambridge CB3 0HA}
\begin{document}

\date{Accepted 2010 December 1.  Received 2010 November 30; in original form 2010 July 9
}

\pagerange{\pageref{firstpage}--\pageref{lastpage}} \pubyear{2010}

\maketitle

\begin{abstract}
We present a study of strong cool core, highly-luminous (most with $\Lx\ge~10^{45}~\ergps$), clusters of galaxies in which the mean jet power of the central active galactic nucleus must be very high yet no central point X-ray source is detected. Using the unique spatial resolution of $Chandra$, a sample of 13 clusters is analysed, including A1835, A2204, and one of the most massive cool core clusters, RXCJ1504.1-0248. All of the central galaxies host a radio source, indicating an active nucleus, and no obvious X-ray point source. For all clusters in the sample, the nucleus has an X-ray bolometric luminosity below 2 per cent of that of the entire cluster. Most have a nucleus $2-10\keV$ X-ray luminosity less than about $10^{42}\ergps$. We investigate how these clusters can have such strong X-ray luminosities, short radiative cooling-times of the inner intracluster gas requiring strong energy feedback to counterbalance that cooling, and yet have such radiatively-inefficient cores. If the central black holes have masses $\sim10^9M_\odot$ then the power exceeds one per cent of the Eddington luminosity, and they are expected to be radiatively-efficient. Only if they are ultramassive ($M_\mathrm{BH}>10^{10}M_\odot$), would their behaviour resemble that of lower mass accreting black holes. Our study focuses on deriving the nucleus X-ray properties of the clusters as defined in the above question, while briefly addressing possible solutions. 
\end{abstract}


\section{Introduction}

Many clusters of galaxies have steeply rising X-ray surface brightness profiles, drops in temperature by a factor of 2 to 3 near the centre \citep{All2001328,deG2002567} and short radiative cooling times. These are known as cool core clusters. With peaked X-ray surface brightness, cool core clusters cool rapidly in their centres. To maintain pressure support, this results in an inflow of matter on the scale of hundreds of solar masses per year, also known as a cooling-flow. However, both $Chandra$ and $XMM-Newton$ show much less cooling gas below a factor of 2 to 3 of the outer temperature than is predicted from the cooling flow model \citep{Boh2001365,Tam2001365,Pet2001365,Pet2003590,Pet2006427,Mcn200745}.

Additional heating is needed to offset the cooling. For cool core clusters, the central active galactic nucleus (AGN) is likely to be the energy source heating the gas by interacting with the ICM \citep{Bir2004607,Raf2006652,Dun2006373,Dun2008385}. 
\begin{figure*}
\centering
\begin{minipage}[c]{0.24\linewidth}
\centering \includegraphics[width=\linewidth]{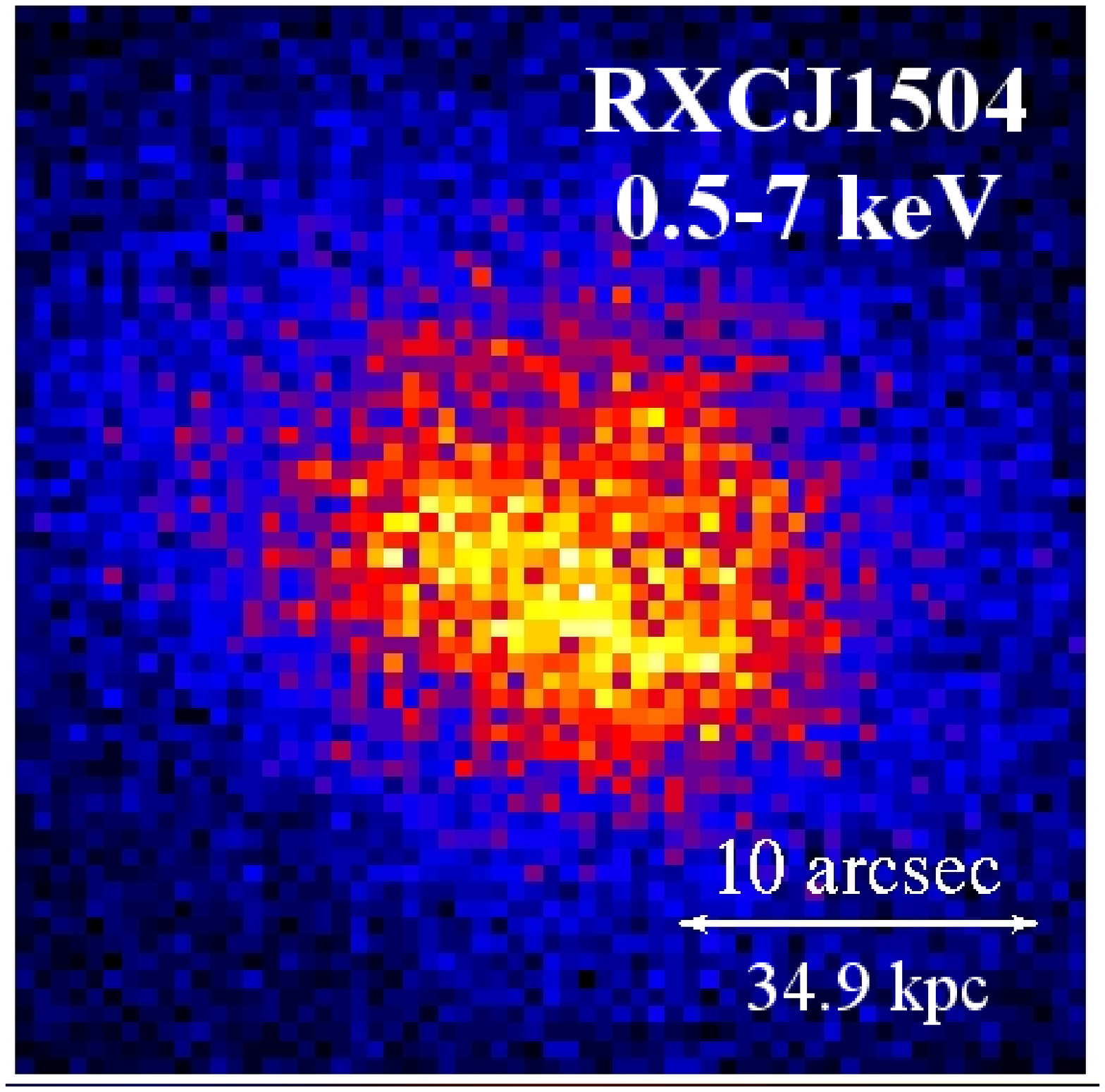}
\end{minipage}
\begin{minipage}[c]{0.24\linewidth}
\hspace{1.5 in}
  \centering \includegraphics[width=\linewidth]{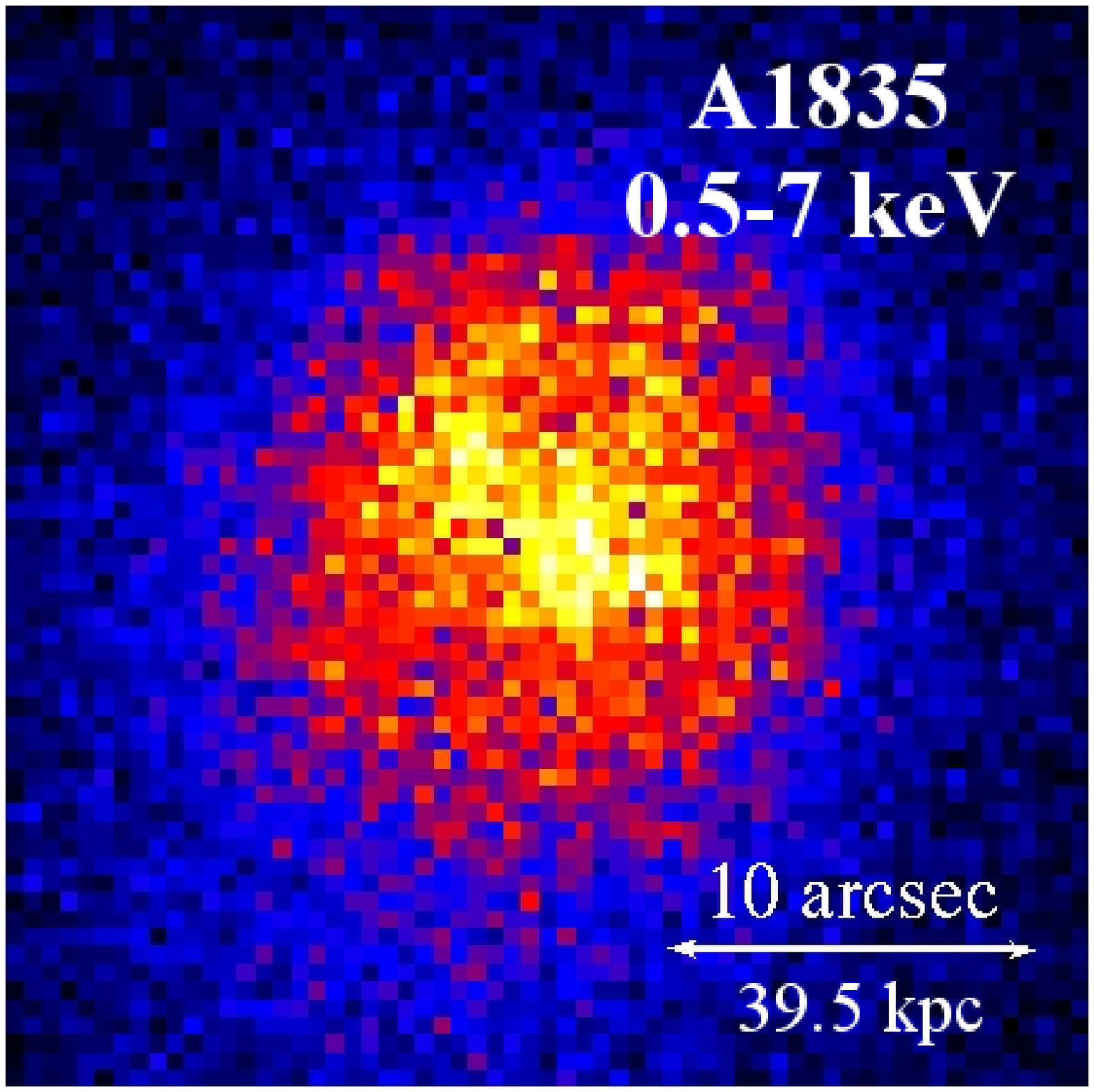}
\end{minipage}
\begin{minipage}[c]{0.24\linewidth}
\hspace{1.5 in}
  \centering \includegraphics[width=\linewidth]{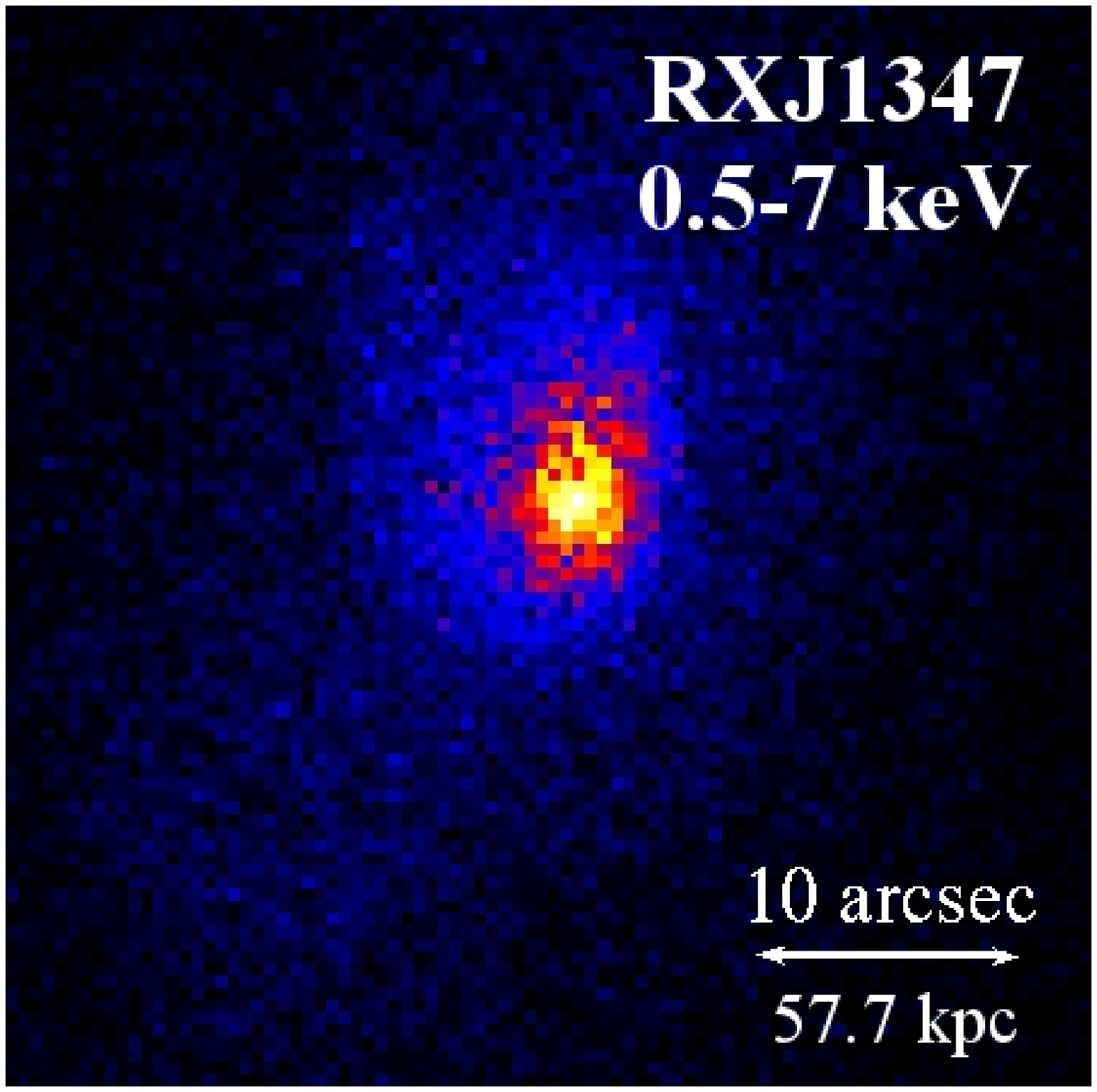}
\end{minipage}
\begin{minipage}[c]{0.24\linewidth}
\hspace{1.5 in}
  \centering \includegraphics[width=\linewidth]{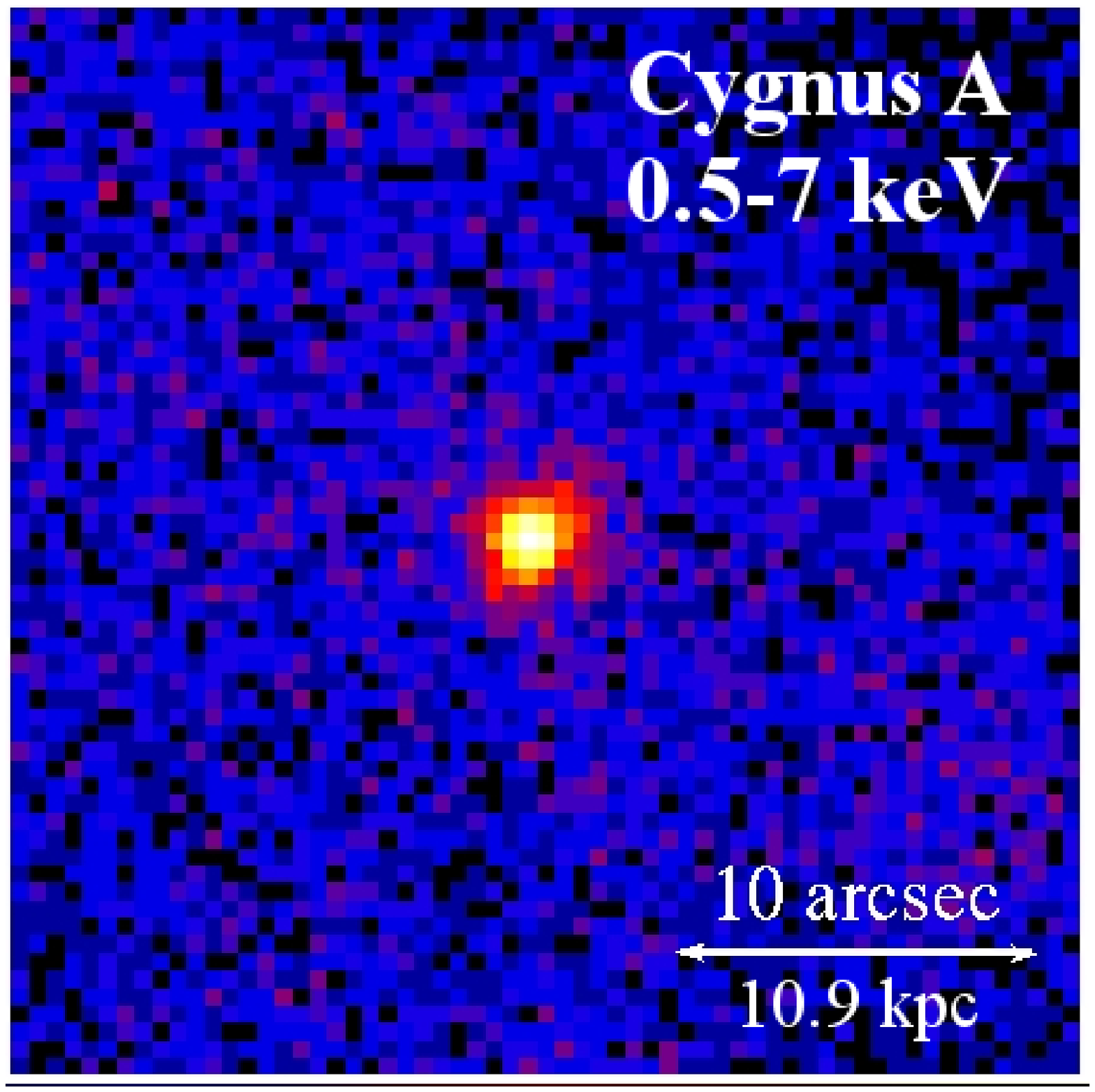}
\end{minipage}
\begin{minipage}[c]{0.24\linewidth}
\centering \includegraphics[width=\linewidth]{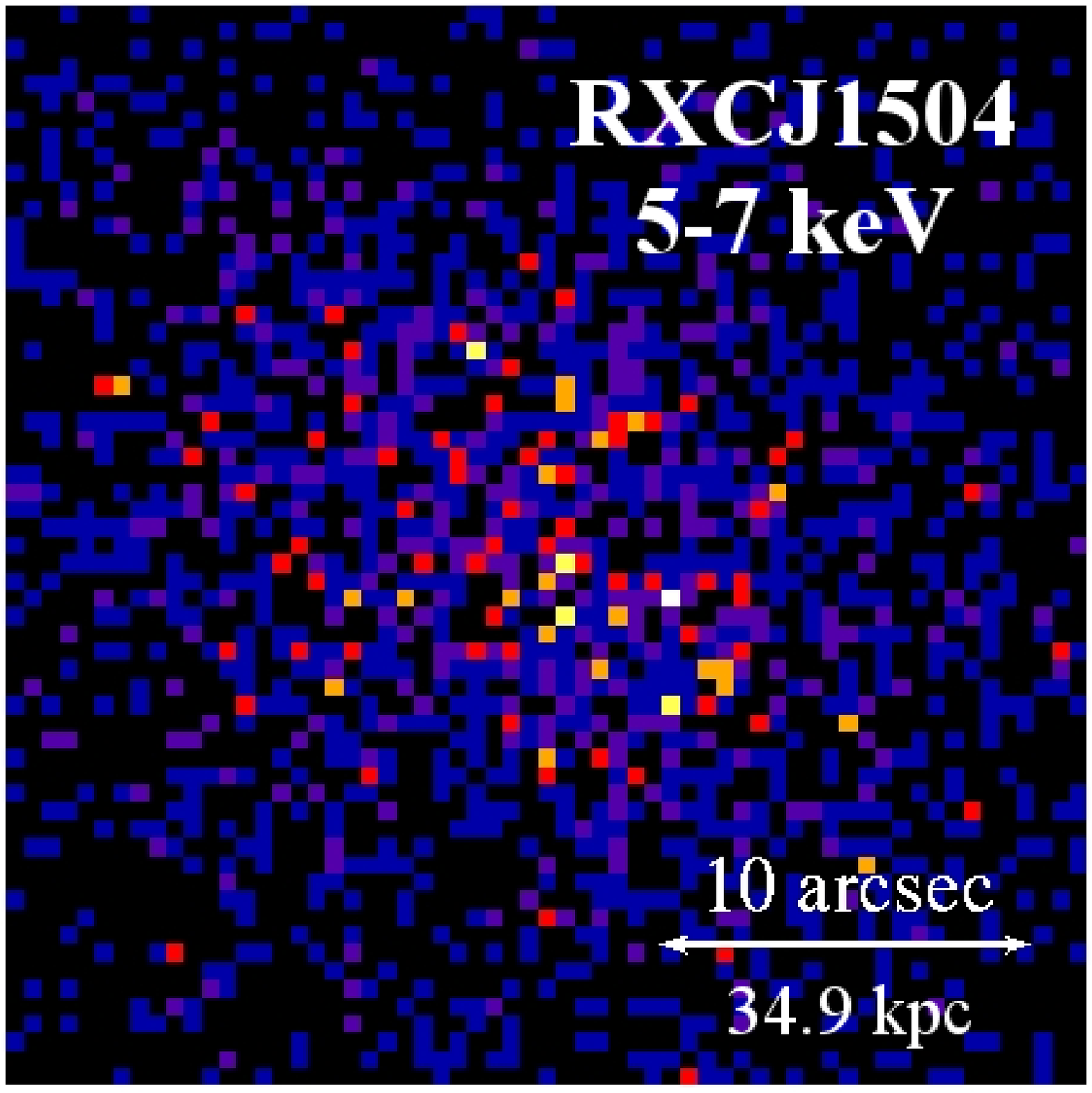}
\end{minipage}
\begin{minipage}[c]{0.24\linewidth}
\hspace{1.5 in}
  \centering \includegraphics[width=\linewidth]{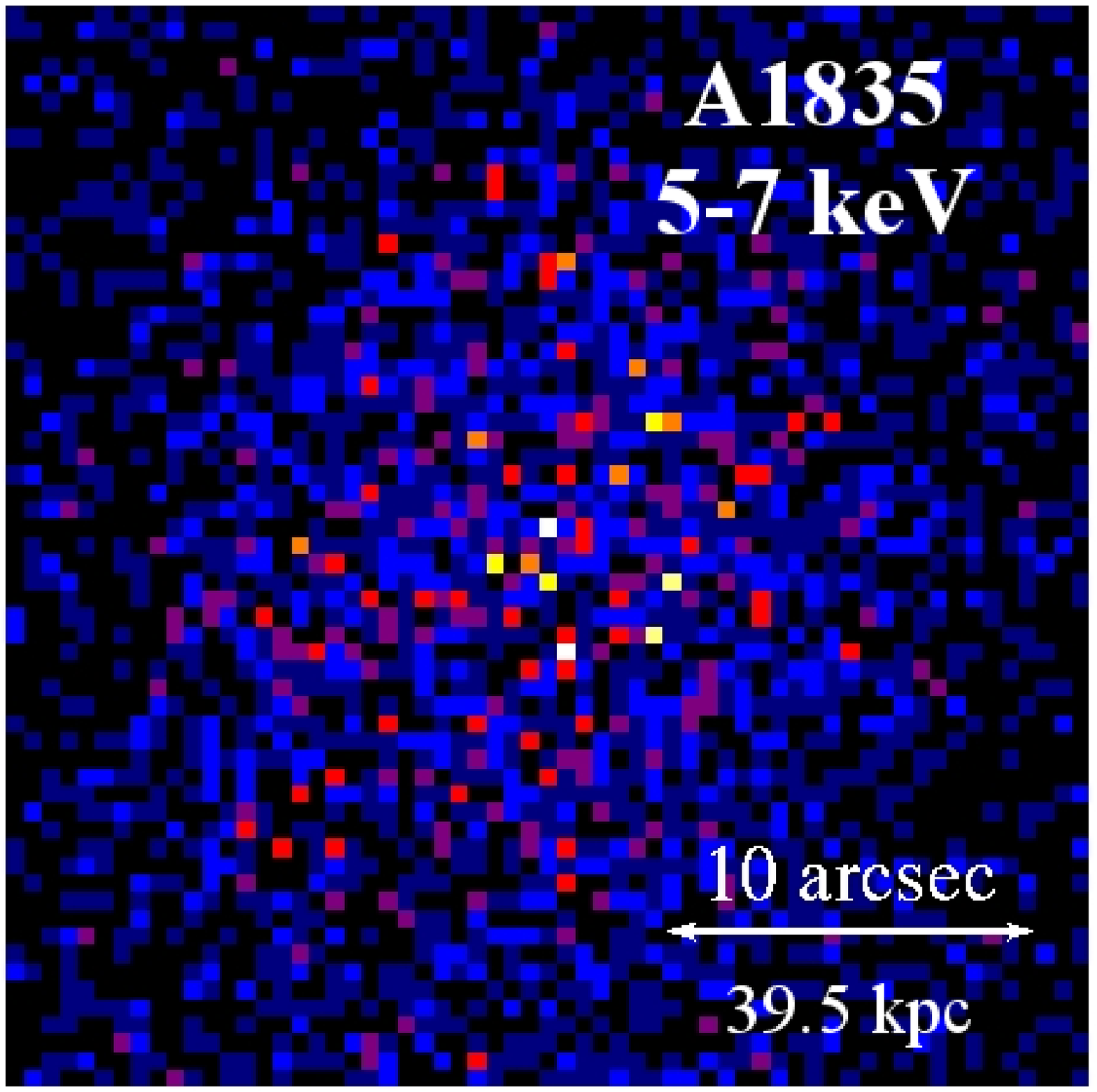}
\end{minipage}
\begin{minipage}[c]{0.24\linewidth}
\hspace{1.5 in}
  \centering \includegraphics[width=\linewidth]{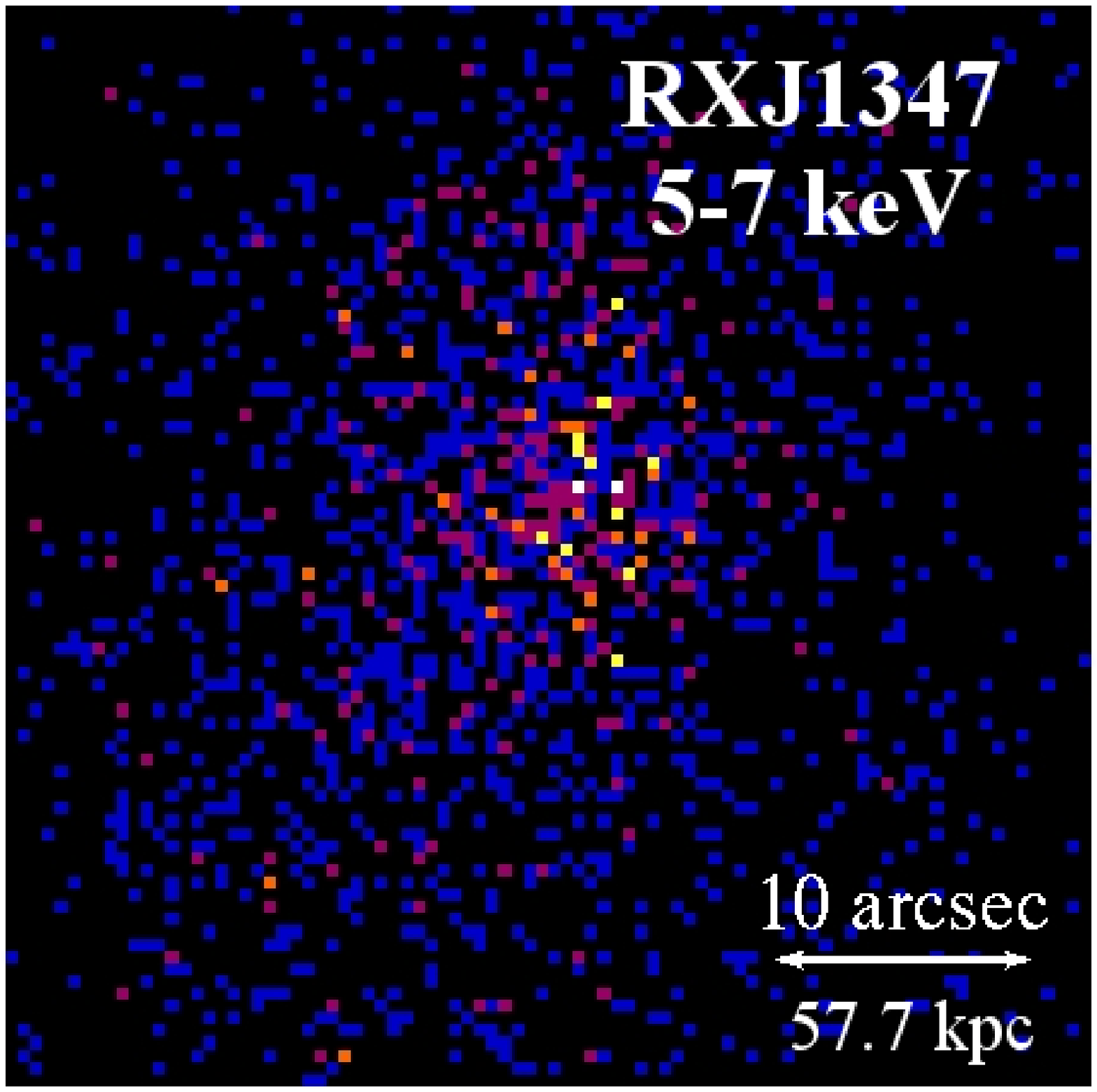}
\end{minipage}
\begin{minipage}[c]{0.24\linewidth}
\hspace{1.5 in}
  \centering \includegraphics[width=\linewidth]{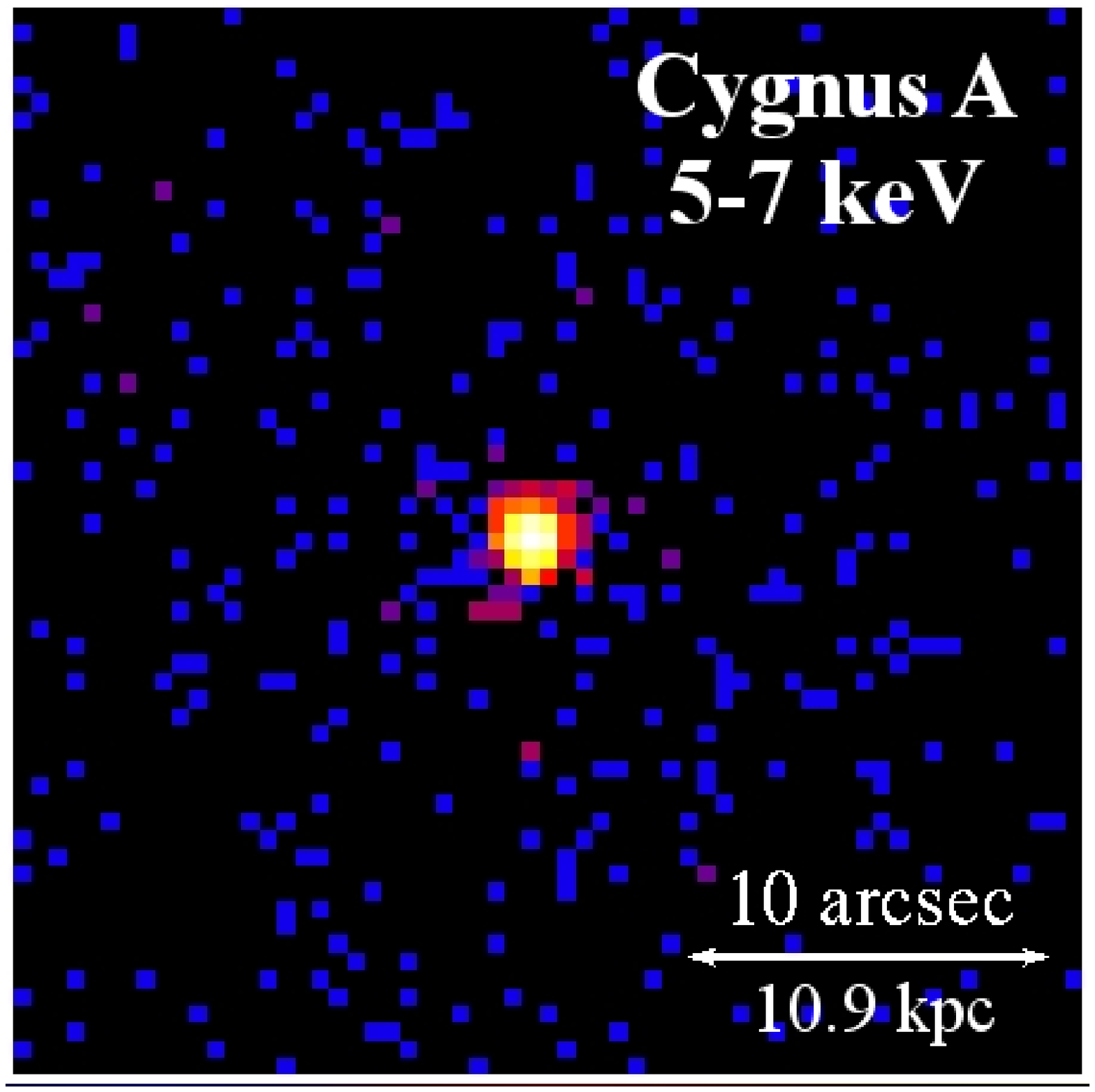}
\end{minipage}
\caption{$Chandra$ images from left to right of R1504, A1835, R1347 and Cygnus A, with one$\pix$ ({\rm e. g.} 0.492$''$) binning. The colour scale is linear expect for Cygnus A which is in a log scale. Black shows regions with zero counts. North is up and East is to the left. The top panels are in the $0.5-7\keV$ energy range, and the bottom panels are in the $5-7\keV$ energy range. The X-ray images of Cygnus A clearly reveal its point-like core, whereas those of R1504 and A1835 do not show any obvious evidence of one. Also shown are the images of R1347 which begin to show a hint of point-like core. }
\label{fig1}
\end{figure*}

Energy from the AGN can be injected into the ICM either radiatively or mechanically. Extreme examples of the first are quasars and bright Seyferts \citep[see][ for an example on how radiation can couple via a wind with the surrounding medium]{Hop2010401}, whereas the second can be seen through outflows and jets in radio galaxies. Cygnus A is an ideal example hosting a very powerful nucleus. $Chandra$ images reveal that this cluster has an obvious point-like core, and a spectral analysis shows that the absorption-corrected core X-ray luminosity accounts for more than 50 per cent of the cluster X-ray luminosity. Radio images of this source reveal large scale jets interacting with the ICM \citep{You2002564}. 

Many authors have shown in recent years that AGN can energetically offset the cooling in galaxy clusters by inflating large cavities correlated with radio lobes, and propagating energy through sound/pressure waves \citep{Fab2003344,Bir2004607,Rus200436,For2005635,Dun2006373,Fab2006366,Raf2006652,Sij2006371,Mcn200745,San2007381,Dun2008385}. However, not all cool core clusters have an obvious point-like core in their X-ray emission like that of Cygnus A. 

\citet{Mer2007381} studied a sample of 15 sub-Eddington AGN for which there was an estimate of the kinetic jet power available from the literature, and a clear nuclear X-ray source reported (although 2 sources only had upper limits of their nuclear source). From there, these authors were able to find a correlation between kinetic power and nuclear X-ray luminosity. Our study presents a sample of strong cool core, highly-luminous ($\Lx\ge10^{45}\ergps$) clusters of galaxies, for which their is $no$ evidence of a nuclear X-ray source in the $Chandra$ images. We select a sample with known, luminous, cool cores from the $Chandra$ archives. It includes some of the most extreme objects such as RXCJ1504.1-0248 \citep[see][]{Ogr2010}. All of the clusters in our sample require very energetic feedback from their AGN to offset cooling, and all have a radio source, indicating that the AGN should be switched on, yet none have an obvious point source in the X-ray band. Unless the central black hole is ultramassive ($M_\mathrm{BH}\gg10^9M_\odot$), our objects are operating at high enough Eddington rates that they should be radiatively efficient. This makes the lack of any evident X-ray nucleus quite puzzling \citep[see also][]{McN2010}. 

In Section 2, we present the sample and reduction techniques. In Section 3, the results are shown and in Section 4, they are discussed. We adopt $H_\mathrm{0}=70\kmpspMpc$ with $\Omm=0.3$, $\OmL=0.7$ throughout this paper.

\section{Sample selection and data analysis}

\begin{sidewaystable*}
\caption{Sample of Clusters - (1) Name; (2) redshift; (3) IR luminosity ($\LIR$); (4) identification number of $Chandra$ observational project; (5) total exposure time; (6) central cooling time in Giga-years; (7) unabsorbed bolometric X-ray luminosity within $r=200\kpc$, with 2$\sigma$ upper and lower uncertainties; (8) cooling radius; (9) kinetic luminosity determined as the unabsorbed X-ray bolometric luminosity within $r=1.34\rcool$, with 2$\sigma$ upper and lower uncertainties; (10) 2$\sigma$ upper limit of background corrected number of counts for the nucleus; (11) and (12) 2$\sigma$ upper limit count rate converted into an unabsorbed nucleus luminosity using {\sc pimms}; (13) and (14) unabsorbed spectroscopic luminosities (rest-frame $2-10\keV$) with an additional adsorption of the power law of $N_\mathrm{H}=0$ (Column 13) and $\mathrm{log}(N_\mathrm{H})=23$ (Column 14) along with the 2$\sigma$ upper and lower uncertainties. When only upper limits were available, the 2$\sigma$ upper limit is shown. The dash symbol (-) indicates either no available data in the literature for Column 3 or no fit to the data for Columns 13 and 14. Comments: $^a$- Galactic absorption was free to vary since the value from \citet{Kal2005440} did not provide a good fit. $^b$- The X-ray emission region for the nucleus was taken as the same one in \citet{Tay2006365}. $^c$- Internal absorption of the power law was always allowed for Cygnus A where we used our spectroscopically derived value ($\mathrm{log}(N_\mathrm{H})=23.33$), but all luminosities shown are corrected for absorption. }
\centering
\resizebox{23.8cm}{!} {
\begin{tabular}{lcccccccccccccc}
\hline
\hline
& & & & & & & & & \multicolumn{3}{c}{\large{PHOTOMETRIC METHOD}} & \multicolumn{2}{c}{\large{SPECTROSCOPIC METHOD}} \\
(1) & (2) & (3) & (4) & (5) & (6) & (7) & (8) & (9) & (10) & (11) & (12) & (13) & (14) \\
Cluster & $z$ & $\LIR$ & Obs ID & Exp. & $\tcool$ & $\Lx$ & $\rcool$ & $\Lkin$ & Counts & $\Lnuc$ & $\Lnuc$ & $\Lnuc$ & $\Lnuc$ \\
& & & & & & & & & 2$\sigma$ limit & 2$\sigma$ limit& 2$\sigma$ limit & $N_\mathrm{H}=0$ & $\mathrm{log}(N_\mathrm{H})=23$\\
& & & & & & $0.01-50\keV$ & & $0.01-50\keV$ & $0.5-7\keV$ & $0.01-50\keV$ & $2-10\keV$ & $2-10\keV$ & $2-10\keV$ \\
 & & ($10^{44}\ergps$) & & ($\rm{ks}$) & ($\rm{Gyr}$) & ($10^{44}\ergps$) & ($\rm{kpc}$) & ($10^{44}\ergps$) & & ($10^{42}\ergps$) & ($10^{42}\ergps$) & ($10^{42}\ergps$) & ($10^{42}\ergps$)\\
\hline

RXJ1532.9+3021 & 0.3450 & 22.6 & 1649 & 9.4 & 0.5 & 29.6$^{+1.1}_{-0.9}$ & 107 & 25.7$^{+0.9}_{-0.9}$ & 80 &  68.2  & 15.1 & $<$11 & $<$16 \\
A1835 & 0.2532 & 28.0 & 6880 & 113.8 & 0.6 & 34.4$^{+0.3}_{-0.4}$ & 96.7 & 26.4$^{+0.2}_{-0.3}$ & 231 & 11.9 & 2.62 & 3.3$^{+0.5}_{-0.7}$ & 6.3$^{+2.2}_{-2.1}$ \\
A2204 & 0.1522 & 3.2 & 7940 & 73.1 & 0.3 & 24.2$^{+0.3}_{-0.2}$ & 69.5 & 15.6$^{+0.1}_{-0.2}$ & 352 & 9.56 & 2.11 & 2.2$^{+0.3}_{-0.6}$ & 3.9$^{+1.4}_{-1.3}$ \\
A1664 & 0.1283 & 3.2 & 7901 & 36.4 & 0.35 & 4.29$^{+0.06}_{-0.06}$ & 61.9 & 2.65$^{+0.04}_{-0.04}$ & 52 & 1.42 & 0.32 & $<$0.24 & $<$0.44 \\
RXCJ1504.1-0248 & 0.2153 & - & 5793 & 34.0 & 0.32 & 50.0$^{+0.6}_{-0.5}$ & 95 & 41.5$^{+0.5}_{-0.5}$ & 160 & 20.1 & 4.45 & 3.4$^{+1.2}_{-1.7}$& 9.4$^{+4.8}_{-4.3}$ \\
RXJ0439.0+0715 & 0.2300 & - & 3583 & 19.0 & 6.1 & 8.1$^{+0.4}_{-0.3}$ & 40.2 & 1.7$^{+0.1}_{-0.1}$ & 21 & 5.62 & 1.24 & $<$1.4 & $<$2.5 \\
A2390 & 0.2280 & 1.2 & 4193 & 77.6 & 1.9 & 21.4$^{+0.4}_{-0.4}$ & 60.9 & 9.1$^{+0.2}_{-0.2}$ & 163 & 6.99 & 1.54 & 1.7$^{+0.2}_{-0.6}$& 3.7$^{+1.8}_{-1.6}$ \\
A0478$^a$ & 0.0881 & - & 1669 & 40.3 & 1.1 & 19.0$^{+0.2}_{-0.2}$ & 84.8 & 12.7$^{+0.1}_{-0.1}$ & 152 & 2.28 & 0.50 & 0.47$^{+0.07}_{-0.31}$& 0.55$^{+0.44}_{-0.34}$ \\
PKS0745-19 & 0.1028 & 3.8 & 2427 & 17.9 & 1.1 & 24.2$^{+0.2}_{-0.2}$ & 89 & 18.3$^{+0.2}_{-0.1}$ & 66 & 3.57 & 0.79 & 0.75$^{+0.30}_{-0.29}$& 1.9$^{+1.3}_{-1.0}$ \\
A2261 &  0.2240 & 0.8 & 5007 & 23.3 & 3.0 & 14.1$^{+0.5}_{-0.5}$ & 51 & 4.6$^{+0.3}_{-0.2}$ & 24 & 4.63 & 1.02 & $<$0.96 & - \\
Z2701 & 0.2151 & $<$1.1 & 3195 & 25.9 & 1.2 & 6.2$^{+0.1}_{-0.2}$ & 67.4 & 3.76$^{+0.1}_{-0.1}$ & 38 & 3.85 & 0.85 & 0.57$^{+0.20}_{-0.43}$& 1.4$^{+1.7}_{-1.0}$ \\
RXJ1720.1+2638 & 0.1640 & $<$1.1& 4361 & 18.4 & 1.9 & 13.2$^{+0.2}_{-0.3}$ & 79.4 & 8.8$^{+0.2}_{-0.2}$ & 54 & 6.60 & 1.46 & $<$2.5 & $<$1.9 \\
RXJ2129.6+0005 & 0.2350 & 2.9 & 9370 & 28.5 & 0.7 & 12.1$^{+0.4}_{-0.3}$ & 72.8 & 7.1$^{+0.2}_{-0.2}$ & 71 & 12.5 & 2.76 & $<$2.2 & $<$3.9 \\
\hline
Z3146 & 0.2906 & 15.7 & 9371 & 36.1 & 0.6 & 30.4$^{+0.5}_{-0.5}$ & 93 & 23.0$^{+0.4}_{-0.4}$ & 188 & 42.2 & 9.31 & 8.2$^{+1.7}_{-4.5}$& 8.4$^{+5.6}_{-4.6}$ \\
MS1455.0+2232 & 0.2578 & 1.1 & 4192 & 78.9 & 0.28 & 16.0$^{+0.1}_{-0.2}$ & 96.6 & 13.8$^{+0.1}_{-0.2}$ & 162 & 12.8 & 2.81 & 2.6$^{+0.6}_{-2.9}$& 3.0$^{+2.4}_{-2.1}$ \\
RXJ1347.5-1145 & 0.4510 & - & 3592 & 51.5 & 2.6 & 100.7$^{+2.9}_{-2.8}$ & 78 & 65.5$^{+2.0}_{-1.8}$ & 358 & 162.3 & 35.9 & $<$48 & 25$^{+19}_{-18}$ \\
MS2137.3-2353 & 0.3130 & 1.3 & 4974 & 7.2 & 1.2 & 18.3$^{+0.8}_{-0.8}$ & 72 & 13.4$^{+0.6}_{-0.6}$ & 61 & 55.8 & 12.3 & 11$^{+2}_{-6}$& 22$^{+19}_{-14}$\\
\hline
Centaurus$^{a,b}$ & 0.0104 & - & 4954 & 87.2 & $<$ 3 & 0.329$^{+0.003}_{-0.003}$ & 42.8 & 0.2639$^{+0.0007}_{-0.0008}$ & 217 & 0.015 & 0.0033 & 0.0017$^{+0.0006}_{-0.0006}$& 0.0029$^{+0.0020}_{-0.0014}$ \\
\hline
Cygnus A$^c$ & 0.0561 & - & 1707 & 9.2 & $<$ 3 & 14.3$^{+1.7}_{-1.8}$ & 55.1 & 13.5$^{+1.4}_{-1.4}$ & 2436 & 1035 & 228.5 & - & 207$^{+17}_{-16}$ \\
\hline
\end{tabular}}
\label{tab1}
\end{sidewaystable*}

We selected a sample of clusters for which luminosities, cooling times and $Chandra$ data were available. From there, we mostly limited ourselves to those that were cool cores ($\tcool\leq7\Gyr$) and highly-luminous with average $\Lx\ge10^{45}\ergps$. The sample contains many of the luminous cool core clusters above a redshift of 0.08, but depends on what is available in the $Chandra$ archive. Our aim was to select a large enough sample to show that our type of object is not rare, but not to select $all$ clusters in the $Chandra$ archive that met with the properties of our objects (i. e. clusters requiring strong feedback from their AGN but with no X-ray detectable nucleus). Our sample is therefore not complete. We suspect that there may be many other clusters that agree with the properties of our objects. 

For our selected sample, the first division of Table~\ref{tab1} outlines the thirteen clusters that did not show any evidence of a point source in their $Chandra$ images. Those in the second (Z3146, MS1455, R1347 and MS2137) show a hint of a point source. Finally, we also included the Centaurus Cluster in order to compare our results with those of \citet{Tay2006365}, as well as Cygnus A, since it is a perfect counter-example\footnote[1]{Other examples of luminous clusters hosting a bright X-ray point source at their centres include H1821+643 \citep[][]{Rus2010402}, IRAS09104+4109 \citep[][]{Iwa2001321}, MACSJ1931.8-2634 \citep[][]{Ehl2010}, 3C295 \citep{All2001324}, 4C+55.16 \citep[][]{Iwa2001328}, and even the Perseus Cluster \citep[e. g. ][]{Fab2000318}.} with an obvious, but faint, point source. In summary, an absence of a detected point source is apparent in about half of the luminous clusters we examine. 

Table \ref{tab1} includes the infrared (IR) luminosities of the BCG in our clusters. These luminosities were taken from \citet{Qui2008176} ($8-1000\mu\rm{m}$), \citet{Ega2006647} ($\sim3-700\mu\rm{m}$) or using the 15$\mum$ luminosity of \citet{Ega2006647} and the equation relating the latter to the IR luminosity of \citet{Elb2002384} ($8-1000\mu\rm{m}$).

Table \ref{tab1} shows the observation identification numbers and exposure times of the X-ray data. We took one of the deepest available non-grating data sets, except for Cygnus~A where we decided to take OBSID 1707 which had a frame rate of only 0.4$\rm{s}$ to avoid pile-up \citep{You2002564}. For those that had data sets with similar exposure times, we selected the most recent one. The cooling times were taken either from Allen (2000), Bauer et al. (2005), Dunn \& Fabian (2006), Rafferty et al. (2008) or Kirkpatrick et al. (2009)\nocite{All2000315,Bau2005359,Dun2006373,Raf2008687,Kir2009697}. The cooling radius was taken from \citet{Dun2006373,Dun2008385}, or as the radius for which the cooling time was equal to 3$\Gyr$ in the online data of \citet{Cav2009218}. For all of our objects, the radius where $\tcool\sim1\Gyr$ is just less than half the listed $\rcool$ or $20-50\kpc$. The region with a short ($<1\Gyr$) cooling time therefore remains large. 

The X-ray $Chandra$ data were processed, cleaned and calibrated using the latest version of the {\sc ciao} software ({\sc ciaov4.2}, {\sc caldb4.2}). Spectra were analysed using {\sc xspec} (v12.6.0d,f)\citep[{\rm e. g.} ][]{Arn1996101}. Fig. \ref{fig1} shows images of four clusters in our sample. The top panels are images in the $0.5-7\keV$ energy range, and the bottom panels are the high energy images ($5-7\keV$) where you would expect to have hints of a point source if a non-thermal component was present.  Fig. \ref{fig1} shows that the X-ray luminosity peaks toward the centre, but that there is no clear evidence of a luminous point-like AGN for R1504 and A1835. 

The cluster luminosities shown in Column 7 of Table \ref{tab1} were calculated by extracting a $0.5-7\keV$ spectrum of the inner 200$\kpc$, binned with a minimum of 30 counts per bin, and then fitting it with a single-temperature {\sc mekal} \citep{Mewe1995} model including {\sc phabs} absorption to account for Galactic absorption. The Galactic absorptions throughout this paper were frozen at the \citet{Kal2005440} values, except for Centaurus and A0478 where the absorption was free to vary in order to provide a good fit. The redshifts were also kept frozen throughout this paper. Using the {\sc cflux} model in {\sc xspec}, we obtained a flux estimate for the cluster corrected for absorption, which was then converted into a luminosity using the luminosity distance. {\sc cflux} provides a better estimate of the flux by exploring the space allowed by the free parameters. The kinetic luminosities shown in Column 9 were calculated in the same way, but within the inner region of a radius ($r$) equal to $1.34\rcool$, where \citet{Dun2008385} found that on average bubbles can offset cooling within this radius. For both luminosities, the 2$\sigma$ upper and lower uncertainties are shown. All error uncertainties were found with the {\sc error} command in {\sc xspec}.  

The flux of the nucleus in each cluster was estimated using two methods, one photometric and another spectroscopic. The first was done using the web interface of {\sc pimms}\footnote[1]{http://heasarc.gsfc.nasa.gov/Tools/w3pimms.html} \citep{Mukai1993}. For each cluster, the location of the nuclear X-ray emission in the $0.5-7\keV$ $Chandra$ images was taken as the 4 by 4 pixels square at the brightest point in the centre of the cluster. The background was taken as a square annulus centred on the same position, with an outer 8 by 8 pixels square and an inner 6 by 6 pixels square. Since the X-ray counts are governed by Poisson statistics, the noise can be estimated as $\sigma=\sqrt{N}$ in each region. The total count number of the nucleus X-ray emission was calculated by subtracting background emission scaled to the same number of pixels. Using the error propagation equation, we then estimated the 2$\sigma$ noise level for the background corrected count number, and determined a 2$\sigma$ upper limit for the number of counts of the X-ray nucleus. This is shown in Column 10 of Table \ref{tab1}. Although background subtracted, this count number could still include some thermal contribution. The luminosities obtained are therefore upper limits for any non-thermal contribution. {\sc pimms} was then used to convert the count rate into a flux and then luminosity in the $0.01-50\keV$ (Column 11) and $2-10\keV$ (Column 12) band. A power law model with photon index of 1.9 was used \citep{Gil2007463}, but our results are not sensitive to our value of the index, at least within $\pm0.2$. For the Centaurus Cluster, we used the same location and region for the X-ray nucleus emission and background as in \citet{Tay2006365}, in order to compare both studies.

We investigated whether we could use the assumption that surface brightness scales with radius as a power law, since the surface brightness scales with the square of the density, and density scales as a power law with the radius \citep{All2006372}. Averaging over each annulus, the count rate per$\pix$ can be used as a measure of the surface brightness profile. The difference between any emission arising from the nucleus and the expected emission arising from a fitted power law to the surface brightness profile should give an estimate of the intrinsic nucleus emission. From this intrinsic emission, {\sc pimms} can then be used to estimate the flux of the nucleus. However, when fitting the surface brightness profiles to the inner $\sim10''$ with a power law, excluding the inner $\sim1''$, this predicted a higher number of counts than that found within the nucleus. Hence, our clusters have surface brightness profiles that flatten within the inner regions, and this extrapolation approach cannot be used to estimate the flux of the nucleus.  

{\sc pimms} gives a first estimate of the luminosities based on a count rate, but it is useful to see if a non-thermal component is hidden within the spectrum of the nuclear X-ray emission. If there is a hidden component, this should provide a more precise estimate of the luminosity. The next two paragraphs describe in detail this method. 

First, we used the same location for the X-ray nucleus as that of the first method, but with a circular region of radius 1$''$. Second, another region was taken as an annulus centred on the same position, but with radii between 2$''$ and 3$''$. Using this annulus as a background, and then fitting a power law model to the data did not provide a good fit. Instead, we used this annulus to estimate the properties of the cluster thermal component, which we then extrapolated to the inner circle of radius 1$''$ (see next paragraph). The background was therefore taken as a region located far from cluster emission. Next, C-statistics were used to account for the low number of counts. However, we used the modified version of C-statistics \citep{Arn}, which allows us to use a background file by loading it in {\sc xspec} with the {\sc back} command. We then fitted the spectrum of the nucleus (circle, $r=1''$) in the $0.5-7\keV$ band with a {\sc mekal} + {\sc power law} model in {\sc xspec}, including Galactic absorption ({\sc phabs}). Additional internal absorption ({\sc zphabs}) of the power-law was allowed. However, the power law component made the model largely unconstrained. Even for the clusters which had long exposure times such as A2204, it was difficult to obtain any constraints on the power law parameters. In order to provide a good fit, it was therefore necessary to freeze most parameters. The next paragraph outlines the parameters that were kept frozen, and explains how the fit was accomplished.  

\begin{figure}
\centering
\begin{minipage}[c]{0.99\linewidth}
\centering \includegraphics[width=\linewidth]{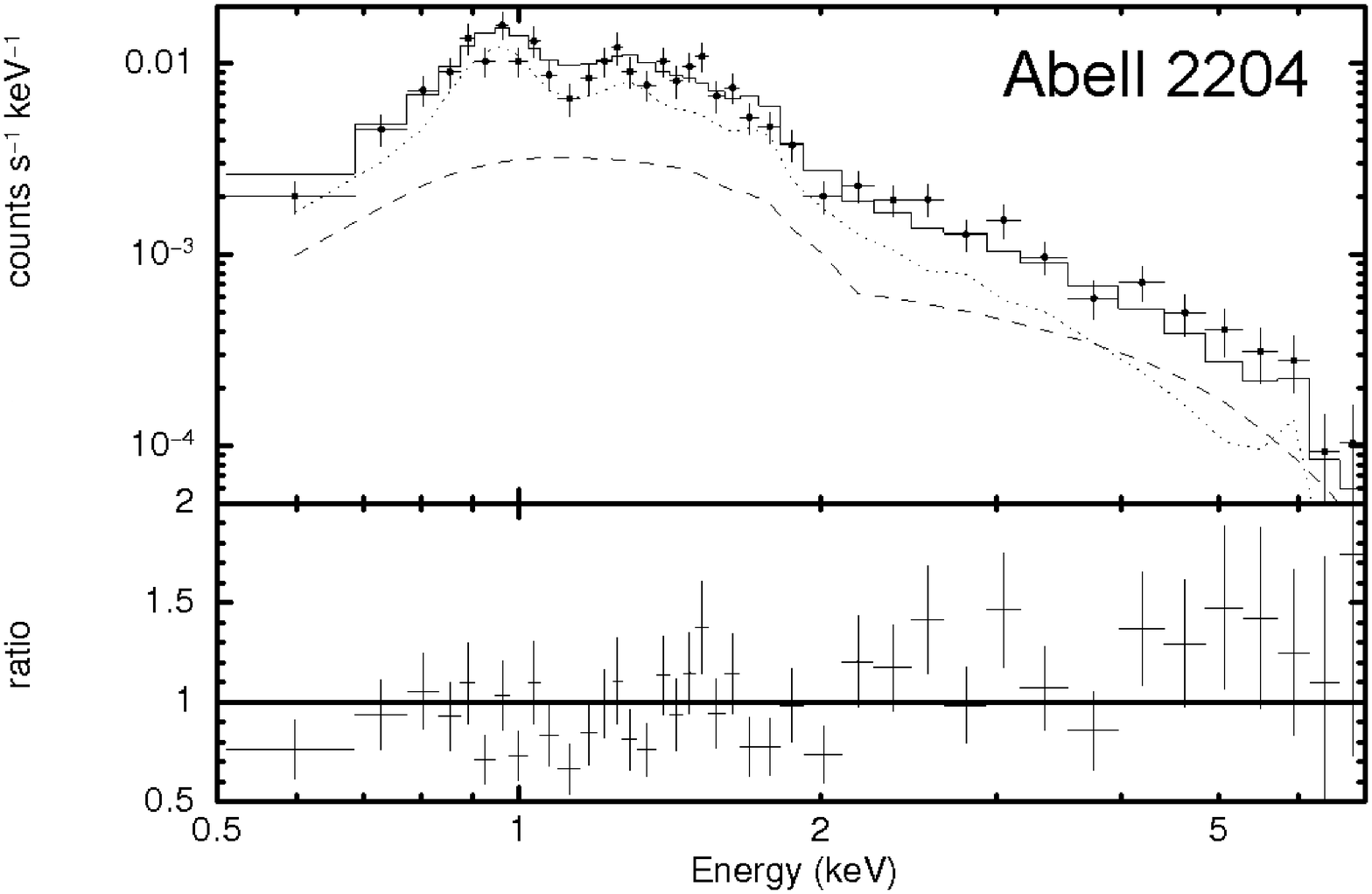}
\end{minipage}
\begin{minipage}[c]{0.99\linewidth}
  \centering \includegraphics[width=\linewidth]{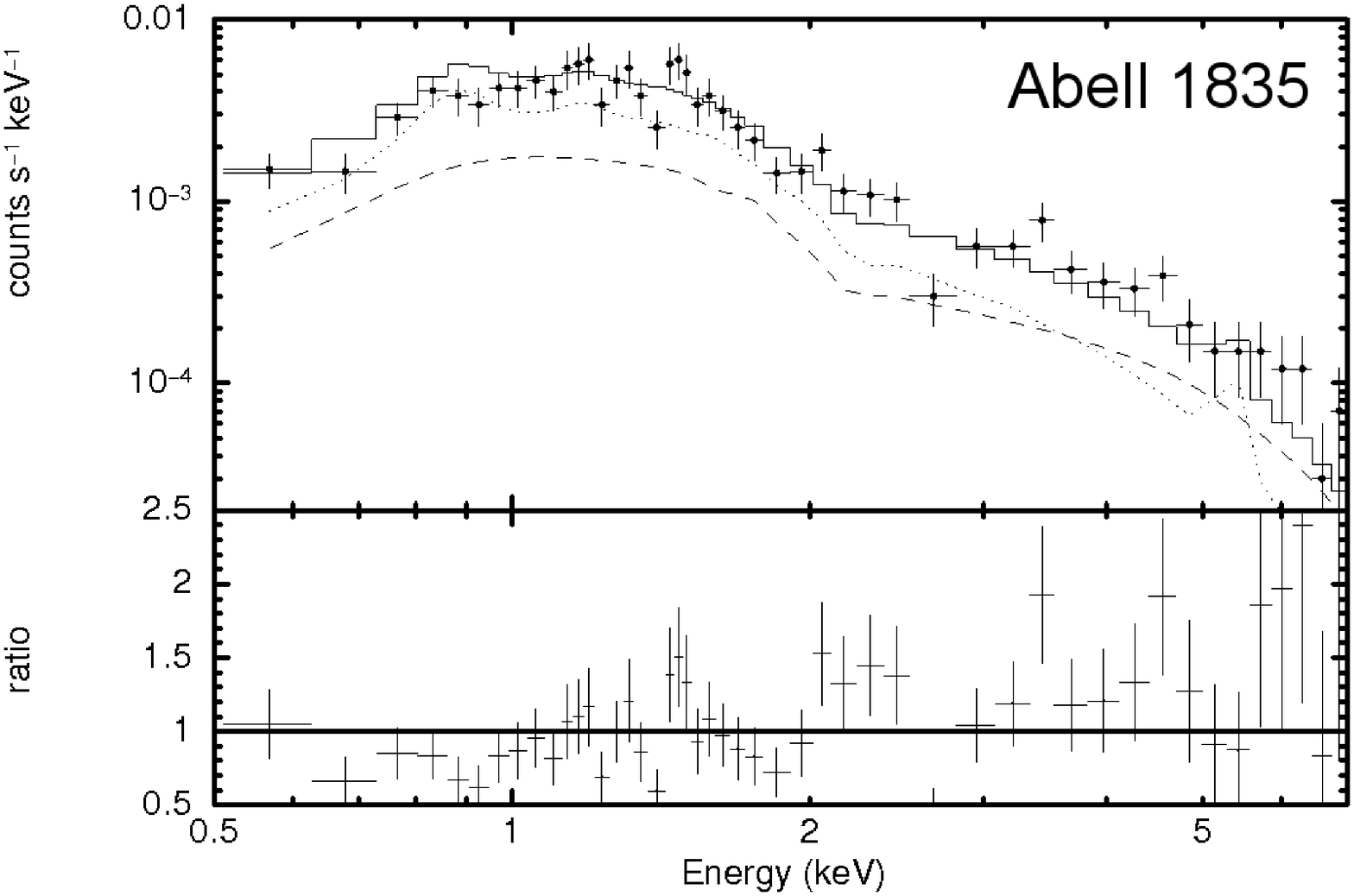} 
\end{minipage}
\caption{Example of one of our good spectral fits of the X-ray nucleus spectrum (A2204, top), and one of our worst fits (A1835, bottom). Each nucleus was fitted with a combined thermal ({\sc mekal}, dotted line) and non-thermal ({\sc pow}, dashed line) model. The combined model is shown with the black line. Galactic absorption was kept frozen at the value of \citet{Kal2005440}, no internal absorption of the power law was allowed, and the background was taken as a region located far from the cluster. The data along with their error bars are shown with the black points, but were rebinned for plotting purposes until a minimum of 5$\sigma$ and a maximum of 30 bins for A2204 and until a minimum of 4$\sigma$ and a maximum of 20 bins for A1835. See Section 2 for more details on the fits.}
\label{fig2}
\end{figure}

First, the Galactic absorption was kept frozen. Next, and as mentioned earlier, the spectrum of the second region (annulus, $r=2-3''$) was used to estimate the properties of the cluster thermal component. We fitted an absorbed (Galactic) single-temperature {\sc mekal} model to the annulus where only the temperature, abundance and normalization were free to vary. The extracted temperature was then extrapolated to the radius of the X-ray emission region (circle, $r=1''$), using the relation that $T=ar^b$ where $b\sim0.3$ \citep{Voi2004347}. The abundance is not expected to vary significantly from $r=1''$ to $r=2-3''$. Using this abundance and extrapolated temperature, along with a power-law index of 1.9, the only parameters allowed to vary in the model fitted to the nucleus emission (circle, $r=1''$) were the normalization parameters of the thermal and non-thermal component. However, the normalization of the thermal component was not allowed to be less than the normalization obtained in the fit for the annulus ($r=2-3''$), scaled for the same pixel number. This is because the normalization parameter is proportional to the square of the density which is expected to increase as the radius decreases. The additional internal absorption for the power-law was kept frozen at either $N_\mathrm{H}=0$ (Table 1, Column 13) or $\mathrm{log}(N_\mathrm{H})=23$ (Table 1, Column 14). See also Fig. \ref{fig2} for some examples. Finally, we extracted the unabsorbed flux of the non-thermal component in the $2-10\keV$ band using the {\sc cflux} model. The 2$\sigma$ upper and lower uncertainties are also shown in Table 1, but some clusters only yielded upper limits, as shown by the $<$ symbol followed by the 2$\sigma$ upper limit. 

We stress that we were able to obtain a rough estimate of the nucleus luminosities based on the spectral analysis, but required most parameters to be frozen. It was not clear whether the addition of the power law model did actually improve the fit. Since our analysis is based on C-statistics, it is not possible to provide a goodness of fit like that of $\chi^2$ statistics. The lack of any obvious non-thermal contribution to the nucleus spectrum of our objects supports the idea that they are radiatively-inefficient.   

X-ray studies of the innermost regions of nearby cool core clusters typically show several cooler components. We strongly suspect that the apparent positive detections of non-thermal components in the present spectral analysis are just manifestations of these cooler thermal components. We therefore proceed by treating all such nucleus detections as upper limits. 

In the case of Cygnus A, the spectrum clearly showed a non-thermal component which required a specific internal absorption. The spectroscopic method was first applied to estimate the luminosity of the nucleus, in the same way as for the other clusters but where the internal absorption of the power law was free to vary. The fitted model gave a value for this absorption of $21.2^{+1.5}_{-1.4}\times10^{22}\cm^{-2}$ ($\mathrm{log}(N_\mathrm{H})=23.33$). Column 14 shows the derived luminosity corrected for absorption. Using this additional absorption, the nucleus luminosity was then estimated using {\sc pimms} in the $0.01-50\keV$ and $2-10\keV$ bands, see respectively in Column 11 and 12. The unabsorbed cluster and kinetic luminosities shown in Column 7 and 9 include the contribution of the non-thermal component of the nucleus. Here, the internal absorption was also kept frozen to the value derived for the nucleus ($\mathrm{log}(N_\mathrm{H})=23.33$). 

Projection effects cannot be corrected for accurately enough near the central regions of clusters, and for this reason we did not consider them throughout this paper. 

\section{Results}

The results are essentially shown in Table \ref{tab1}. \citet{Tay2006365} found a 3$\sigma$ upper limit of $1.2\times10^{40}\ergps$ for the nucleus of Centaurus. They used the annulus surrounding the nucleus as a background and by combining various data sets (200$\ks$) were able to constrain more parameters including the internal absorption of the power law ($N_\mathrm{H}\sim3\times10^{22}\pcmsq$). We compare this with our {\sc pimms} method which also uses the surrounding annulus as a background. Including the same absorption and our 2$\sigma$ upper limit count rate, we derive an unabsorbed upper limit of $1.56\times10^{40}\ergps$ for the nucleus, consistent with \citet{Tay2006365}. 

For Cygnus A, we find an internal absorption of $21.2^{+1.5}_{-1.4}\times10^{22}\cm^{-2}$ and a $2-10\keV$ unabsorbed luminosity for the nucleus of $2.1^{+0.2}_{-0.2}\times10^{44}\ergps$, using a power-law index of 1.9. For the same object, \citet{Uen1994431} found an internal absorption of $37.5^{+7.5}_{-7.1}\times10^{22}\cm^{-2}$ and a $2-10\keV$ unabsorbed luminosity of $\sim5.5\times10^{44}\ergps$ (using our cosmology). Their best-fit power-law index was $1.98^{+0.18}_{-0.20}$, and they estimate at the 90 per cent confidence level that their intrinsic luminosities are uncertain by a factor of 2. Within this uncertainty, our results are consistent with theirs. Our results are also consistent with those of \citet{You2002564}. These authors used the same $Chandra$ data as our study, and found a photon index of $\sim1.5$, an internal absorption of $20^{+1}_{-2}\times10^{22}\cm^{-2}$ and a $2-10\keV$ unabsorbed luminosity of $\sim1.9\times10^{44}\ergps$ (using our cosmology). 

\begin{figure}
\centering
\begin{minipage}[c]{0.99\linewidth}
\centering \includegraphics[width=\linewidth]{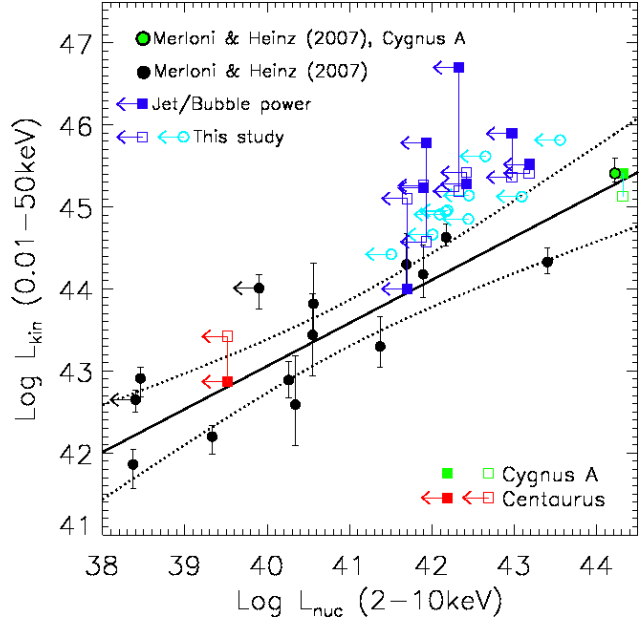}
\end{minipage}
\caption{Kinetic bolometric luminosity versus nuclear luminosity in the $2-10\keV$ band. The AGN of Merloni \& Heinz (2007) are shown with black filled circles. Our results are shown with the non-filled symbols, where our value for $\Lcool(r\le1.34\rcool)$ is used as the kinetic luminosity and the upper limit {\sc pimms} estimate of the $2-10\keV$ nucleus luminosity is used for $\Lnuc$. For Cygnus A, since its power law component is clearly detected, we used our derived spectroscopic value of $\Lnuc$. The green symbols are for Cygnus A and the red symbols are for Centaurus Cluster. The clusters that have estimates of their kinetic luminosity from measurements of the energy output from bubbles/jets are shown in squares and the estimate is shown with the filled square. The black line shows the linear fit to the data, and the dotted lines are the 2$\sigma$ limits.}
\label{fig3}
\end{figure}

In order to illustrate the power needed to offset cooling and the radiative power of the AGN, as well as compare our results with those of \citet{Mer2007381}, we plot the kinetic power versus the nuclear luminosity in the $2-10\keV$ band (Fig. \ref{fig3}). Here the results of \citet{Mer2007381} are shown with black filled circles. Using the partial Kendall's $\mathrm{\tau}$ correlation test, they found that the kinetic power was correlated with the nuclear X-ray luminosity to more than a 3$\sigma$ level. In order to guide the eye, we plot in Fig. \ref{fig3} the linear regression to these data in a log-log plot with a black line, along as the 2$\sigma$ upper and lower limits (dotted lines). The linear regression consisted of minimising the $\chi^2$, and the error bar estimates of $\Lkin$ were taken as the mean value between the upper and lower error bar measurements. \citet{Mer2007381}'s data included two objects which only had upper limits of $\Lnuc$. Our linear regression does not include error estimates of $\Lnuc$. We therefore omitted these two objects for the regression. The non-filled symbols in Fig. \ref{fig3} illustrate our results. Here, the kinetic luminosities and upper limit of the nucleus luminosity are taken from Table \ref{tab1}, the later using the upper limit values derived with {\sc pimms}. For Cygnus A we used the spectroscopic derived value of $\Lnuc$, since it is clearly detected. The clusters that have square symbols also have detailed estimates of their jet power, obtained from observed bubbles. For R1532, A1835, A2204, A0478, P0745, Z2701 and Z3146, this estimate is shown with the dark blue filled square and was taken in order of preference from \citet{Raf2006652,Dun2008385,San2009393}. Cygnus A is shown in green, as well as its jet power estimate in filled green taken from \citet{Mer2007381}. Centaurus is shown in red and its jet power in filled red taken from \citet{Raf2006652}. 

This figure shows that our sample of clusters mostly lie well above the relation of \citet{Mer2007381}'s data, but that Cygnus A and Centaurus fall well inside the confidence levels. All of the clusters in our sample that do not have an obvious X-ray point source have jet powers above $10^{44.4}\ergps$. Hence, they seem to require a jet power from their AGN that is very high. On the other hand, if our objects are simply shifted to the left because they are highly obscured, they would require a large intrinsic absorption of about $\mathrm{log}(N_\mathrm{H})=23.6$. 

\section{Discussion}

Our results reveal a significant population of objects requiring high kinetic input from an AGN to offset cooling and/or high jet power, yet are without a detected X-ray nucleus. These objects appear to be radiatively inefficient with, on average, $\Lkin$/$\Lnuc$ exceeding 200. Our clusters have radio luminosities at 1.4$\GHz$ of about $10^{40}-10^{41}\ergps$, but require very strong jet powers to offset cooling ($>~10^{44.4}\ergps$). When comparing these results to \citet{Bir2008686}'s relation between jet (cavity) power and $\LR$ at 1.4$\GHz$, we find that our clusters lie above this relation, implying once more than our objects require extreme kinetic feedback.   

\subsection{Stellar mass black holes}

The upper-limits to the X-ray luminosity of the nucleus in our sample are at least 10 times lower than is straight forwardly predicted from the properties of the detected lower jet power objects discussed by \citet{Mer2007381} (see Fig. \ref{fig3}). Stellar mass black hole systems may have similarly high ratios of kinetic to luminous power, such as Cygnus X-1 which is thought to spend most of its time (90 per cent) in a low state, with a hard spectrum and collimated jet. \citet{Gal2005436} found that the jet in Cygnus X-1 appears to be inflating a large ring-like structure, and that the kinetic power required can be as high as the bolometric X-ray luminosity of the system. These authors then suggested that, in general, stellar mass black holes dissipate most of their accreted power into radiatively-inefficient relativistic outflows, and not in the form of X-ray emitting inflow. A similar conclusion was reached by \citet[][]{Kor2006369} but for BHs also found in AGN, {\rm i.e.} that a BH in a hard state looses most of its energy in jet power. \citet[][]{Kor2008383} then found that all low-luminosity AGN seem to launch strong jets. 

We do not know the black hole (BH) mass for most of our objects but we expect that they lie between $10^9$ and $10^{10}\Msun$, based on other Brightest Cluster Galaxies (BCGs). Given that these objects are in an exceptional environment and are active continuously, it is not clear that the standard $M_\mathrm{BH}-\sigma$ or $M_\mathrm{BH}-M_\mathrm{K}$ relations are relevant for them. The few BCGs with a reliable $M_\mathrm{BH}$ ({\rm e. g.} M87, Macchetto et al. 1997, as well as A3565, A1836 and A2052, Dalla Bont{\`a} et al. 2009)\nocite{Mac1997489,Dal2009690} indicate the above mass range. The kinetic power then corresponds to 1 to 0.1 per cent of the Eddington limit.

\citet{Chu2005363} have argued, by analogy with the interaction of X-ray binaries, that black holes operating at powers close to the Eddington limit ($L_{\rm Edd}$) are radiatively efficient while at low power (less than 1 per cent $L_{\rm Edd}$), their efficiency drops steadily to low values. The bulk of the power is increasingly taken up by outflows at low Eddington rates. This explains the behaviour of the low and intermediate luminosity galaxies, groups and clusters but is difficult to accommodate for the highest luminosity objects found here \citep[see also][]{McN2010}. The power emerging from our objects means that they are operating above $10^{-2}L_{\rm Edd}$ if $M_\mathrm{BH}\sim10^9M_\odot$. The radiative power should then be at least comparable to the kinetic power. However, if the central black holes significantly exceed $10^{10}M_\odot$, {\rm i. e.} are ultramassive, then the radiative power could fall well below the kinetic power, which is conceivable for our powerful objects but remains untestable at present.

\subsection{Relativistic and geometric effects}

Among the possible explanations as to why our objects have such low observed X-ray luminosities, we first mention that they might simply be Doppler-suppressed. If the X-rays are mostly produced in the jet, and if the jet axis of the AGN lies close to the plane of the sky, as well as if the jet has a high Lorentz factor, the core luminosity would be suppressed by a significant factor such that $\Lnuc=D^3\times\Lnuc_{\rm{, 0}}$. Here, $\Lnuc$ is the observed value of the core luminosity, $\Lnuc_{\rm{, 0}}$ is the intrinsic luminosity and $D$ is is the Doppler factor given by $D=1/(\Gamma(1-\beta\times\cos\theta))$ ($\Gamma$ is the Lorentz factor and $\theta$ is the angle between the line of sight and the jet axis). However, in order for our objects to sit on the relation of \citet{Mer2007381}, their jet axis would not only have to lie close to the plane of the sky, but the jet velocity would also have to be such that $\beta\ge0.9$. It would be unlikely that all of our objects have jets with a preferred geometry (axis aligned with the plane of the sky), and since our sample is not statistically complete, it is possible that there are other objects that meet the properties of those in our sample.

\subsection{Advection dominated accretion flows or magnetically-dominated black holes}

Second, we mention that they might be standard advection dominated accretion flows (ADAFs). Although ADAFs models can create strong winds and relativistic jets, it is not certain that they would be able to create the extreme relativistic jet powers required to inflate the bubbles in our objects, since ADAFs lose much matter in their winds \citep{Nar200851}.

We also consider whether they are powered by magnetically-dominated accretion \citep[{\rm e. g.} ][]{Bla1982199}. Our objects have so far shown no evidence for non-relativistic winds. They may instead be simply magnetically-dominated systems (more than 90 per cent) harbouring powerful jetted outflows. \citet{Kun2007311} and \citet{Mer2002332} pointed out that energy and angular momentum could be removed from an accretion disc in the form of powerful Poynting-dominated outflows if strong magnetic fields were present, while being radiatively-inefficient. The mechanism responsible for creating jets in magnetically-dominated accretion discs still remains poorly understood, which makes it difficult to give any definitive conclusion. 

\subsection{Highly-absorbed AGN}

On the other hand, X-ray absorption can render the nucleus undetectable in $0.5-7\keV$ X-rays. The power must then emerge at longer wavelengths. \citet{Ega2006647} studied $Spitzer$ data of a sample of eleven brightest galaxies in X-ray luminous clusters. This sample included Z3146, A1835, MS1455, A2261, MS2137 and A2390. Essentially, they found that these clusters, which had the shortest cooling times in their sample, had the largest BCG IR luminosities. This seems to suggest that highly-luminous strong cool core clusters have BCGs with larger IR luminosities. These authors also found that A1835 and Z3146 could be classified as LIRGs. Large IR luminosities could be a sign of strong absorption, which would render the X-ray nucleus almost invisible. 

\citet[][]{Gan2009502} reported that there was a 1:1 correlation between the intrinsic $2-10\keV$ X-ray luminosity of an AGN ($\Lnuc$) and its mid-IR luminosity at $\sim12\mu{\rm m}$, implying that both quantifies are intrinsically related \citep[see also][ who looked at a more complete sample of radio sources]{Har2009396}. If we assume that our objects must follow this relation (although they lie at the centre of some of the most extreme cool core clusters), then those that have large mid-IR luminosities (i.e. R1532, A1835, and Z3146) must have large intrinsic nuclear X-ray luminosities on the order of $10^{44}\ergps$. This would mean that they are highly obscured, since they are almost invisible in our $Chandra$ data with $\Lnuc(2-10)\keV\sim10^{42-43}\ergps$. The intrinsic nuclear X-ray luminosities, as predicted from the X-ray/mid-IR relation, would also make them consistent with \citet{Mer2007381}'s relation (see Fig. \ref{fig3}). For the remaining objects in our sample, if they are required to follow the X-ray/mid-IR relation, then this would imply that they are not highly obscured, since the predicted $\Lnuc$ would be consistent with those that we find (i. e. $\sim10^{42}\ergps$). It remains that with the advent of $ALMA$ (Atacama Large Millimeter/submillimeter Array), we will be able to resolve in detail the dusty structures within the core of clusters through sub-mm observations, which will allow us to confirm if the large IR luminosities of some of our BCGs are of nuclear origin. We must stress, however, that most of the far IR emission in our objects is likely due to young stars. There is much evidence from excess blue light, that there is ongoing star formation in these objects \citep[see ][]{ODe2008681}. Nevertheless, there is still room for a significant AGN contribution.

If our objects have different dust properties than those that follow the X-ray/mid-IR relation found by \citet[][]{Gan2009502}, they could still be heavily absorbed. In Column 14 of Table \ref{tab1}, we have shown that even with a $\mathrm{log}(N_\mathrm{H})=23$, resembling the absorption of Cygnus A, the X-ray nucleus luminosity is at most 3 times higher than without internal absorption. If our objects are very heavily absorbed ($\mathrm{log}(N_\mathrm{H})=24$), this would make their non-thermal component almost invisible in the $0.5-7\keV$ X-ray energy band. However, our objects would still represent a different population from Cygnus A which has an obvious point-like core accounting for more than 50 per cent of $\Lx$ even with strong absorption ($\mathrm{log}(N_\mathrm{H})=23.33$). With today's high energy X-ray telescopes, it remains difficult to properly image the high energy band, since the background, sensitivity and spatial resolution remain problematic. However, in the near future, satellites such as $NuSTAR$ (nuclear spectroscopic telescope array, $6-79\keV$) and $Astro-H$ ($0.3-80\keV$) could provide the sensitivity and resolution needed to analyse the power-law component of our objects at high X-ray energy.  

\subsection{AGN duty cycles}

Fifth, we consider strong variability. The duty cycle of an AGN, which measures the probability that a BH is in an active phase is less than 10 per cent \citep[\rm e. g. ][]{Sha2009690}. For our objects, this requires that the power be above $10^{46}\ergps$ when the AGN is switched on. If the duty cycle is less than 1 per cent \citep[{\rm e. g.} ][]{Cio2007665}, the power must exceed $10^{47}\ergps$, which is greater than the most luminous jets known and would surely disrupt the core. Such observed high power jets are not associated with BCGs, but rather blazars. The only quasar in a BCG below $z=0.5$ is H1821+643 \citep{Rus2010402}. In other words, if this explanation is correct, then contrary to observations there should be a population of powerful quasars and blazars at low redshifts associated with BCGs. However, \citet{Dun2008385} found that out of a sample of 34 clusters, only 6 had extreme outbursts capable of offsetting cooling to more than twice the cooling radius. Of these, Z3146 and Z2701 were included. This implies that the duty cycle of extreme outbursts is about 18 per cent, which would require that the power be about $5\times10^{45}\ergps$ when the AGN in our sample are in this mode. 

On the other hand, it is also possible that the properties of our objects are not related to the variability of the jet itself, but rather the variability of how many X-rays are produced by the core, {\rm i.e.} the radiative efficiency of the jets. Our objects may have jets that are continuously pumping $10^{45}\ergps$ into the medium (which would provide the necessary amount of energy needed to heat the ICM), but are in a very radiatively inefficient mode. A scenario here could be to assume that the radiate efficiency of a jet depends on its environment, {\rm i.e.} depends on just how much interaction directly occurs between the jet and surrounding gas. Extremely powerful steady jets may create a channel which results in little such interaction for the bulk of the jet.
 
 \subsection{Spin-powered black holes}
 
Finally, our black holes could also be powered by spin, not accretion \citep[{\rm e. g.} ][]{Beg198456}. \citet{McN2009698} studied the case of MS0745.6+7421 harbouring one of the most powerful outbursts ($\sim10^{62}\erg$) but which has no X-ray point-like core in the $Chandra$ image. They suggest that the outburst could be powered by angular momentum originating from a rapidly rotating BH \citep[see also ][]{McN2010}. Strong jets could be powered by rotating BHs (see {\rm e. g.} the BZ mechanism of Blandford $\&$ Znajek 1977\nocite{Bla1977179}, but see also Fender, Gallo $\&$ Russell 2010\nocite{Fen2010} for a discussion on the observed lack of any correlation between jettedness and spin in galactic microquasars). To operate a spin-powered feedback loop and maintain the balance between cooling and heating, the energy must be continuously tapped. A high spin parameter is required, which according to our significant population of objects would require that highly rotating BHs are not rare. In order to test this theory indirectly, one could look at the relation found by \citet{All2006372}. In their sample of nearby ellipticals, which includes galaxies with an obvious point-like X-ray nucleus, the Bondi accretion rate correlates with jet power. If the spin explanation is correct, then this correlation would not hold for all objects.

\section{Summary}

We have identified a population of objects which require powerful jets to be present but have no X-ray detectable nucleus. We have also identified a range of possible explanations, each of which carries significant implication for the origin and operation of jets. The black holes may be ultramassive ($M_\mathrm{BH}\gg10^9M_\odot$), or have very high spin, or be highly obscured. They may also be mostly off, yet unobservable when they are switched on, or have highly radiatively inefficient jets.

\section*{Acknowledgments}
We thank Helen Russell and Jeremy Sanders for helpful discussions, as well as the latter for providing the internal absorption of the power law derived for Centaurus in \citet{Tay2006365}. JHL also recognizes all the support given by the Cambridge Trusts, Natural Sciences and Engineering Research Council of Canada (NSERC), as well as the Fonds Quebecois de la Recherche sur la Nature et les Technologies (FQRNT). ACF thanks the Royal Society for support. 

\label{lastpage}
\bibliographystyle{mn2e}
\bibliography{bibli}
\end{document}